\documentclass[floats,prd,aps,showpacs,amssymb,nofootinbib,onecolumn]{revtex4}
\usepackage{amssymb}
\usepackage{graphics}
\usepackage{amsmath}
\usepackage{amsfonts}
\usepackage{bm}
\usepackage[]{latexsym}

\newcommand{\be}{\begin{equation}}\newcommand{\ee}{\end{equation}}
\newcommand{\bea}{\begin{eqnarray}}\newcommand{\eea}{\end{eqnarray}}
\newcommand{\brr}{\begin{array}}\newcommand{\err}{\end{array}}
\newcommand{\bit}{\begin{itemize}}\newcommand{\eit}{\end{itemize}}
\newcommand{\ben}{\begin{enumerate}}\newcommand{\een}{\end{enumerate}}

\newcommand{\ba}{\begin{array}}
\newcommand{\ea}{\end{array}}

\def\al{\alpha}

\def\1{{_{1}}}\def\2{{_{2}}}

\def\noHe0{:\;\!\!\;\!\!:H_e(0):\;\!\!\;\!\!:}
\def\noHm0{:\;\!\!\;\!\!:H_\mu(0):\;\!\!\;\!\!:}
\def\al{\alpha}

\def\1{{_{1}}}\def\2{{_{2}}}

\begin{document}
\title{Role of neutrino mixing in accelerated proton decay}

\author{M.Blasone\footnote{blasone@sa.infn.it}$^{\hspace{0.3mm}1,2}$, G.Lambiase\footnote{lambiase@sa.infn.it}$^{\hspace{0.3mm}1,2}$, G.G.Luciano\footnote{gluciano@sa.infn.it}$^{\hspace{0.3mm}1,2}$ 
and L.Petruzziello\footnote{lpetruzziello@na.infn.it}$^{\hspace{0.3mm}1,2}$} \affiliation
{$^1$Dipartimento di Fisica, Universit\`a di Salerno, Via Giovanni Paolo II, 132 I-84084 Fisciano (SA), Italy.\\ $^2$INFN, Sezione di Napoli, Gruppo collegato di Salerno, Italy.}

\date{\today}
\def\be{\begin{equation}}
\def\ee{\end{equation}}
\def\al{\alpha}
\def\bea{\begin{eqnarray}}
\def\eea{\end{eqnarray}}

\begin{abstract}
The decay of accelerated protons has been analyzed both in the laboratory frame (where the proton is accelerated) and in the comoving frame (where the proton is at rest and interacts with the Fulling-Davies-Unruh thermal bath of electrons and neutrinos). The equality between the two rates has been exhibited as an evidence of the necessity of Fulling-Davies-Unruh effect for the consistency of Quantum Field Theory formalism.  Recently, it has been argued that neutrino mixing can spoil such a result, potentially opening new scenarios in neutrino physics. In the present paper, we analyze in detail this problem and we  find that, assuming flavor neutrinos to be fundamental and working within a certain approximation, the agreement can be restored. 
\end{abstract}

\vskip -1.0 truecm 

\maketitle

\section{Introduction}
\setcounter{equation}{0}
It was pointed out by M\"{u}ller~\cite{Muller:1997rt} that the decay properties of particles can be changed by acceleration. In particular, it was shown that usually forbidden processes such as the decay of the proton
become kinematically possible under the
influence of acceleration, thus leading to a finite lifetime for even supposedly
stable particles.
Drawing inspiration from this idea,  Matsas and Vanzella~\cite{Matsas:1999jx,Matsas:1999jx2,Matsas:1999jx3} analyzed the decay of uniformly accelerated protons both in the laboratory and comoving frame, showing that the two rates perfectly agree only when one considers Minkowski vacuum to appear as a thermal bath of neutrinos and electrons for the accelerated observer (comoving frame). This has been exhibited\footnote{Similar arguments apply, for example, to the analysis of the QED bremsstrahlung radiation. In this case, it has been shown that the emission of a photon by accelerated charges in the inertial frame can be seen  as either
the emission or the absorption of a zero-energy photon in the FDU thermal bath of the comoving observer~\cite{Higuchi:1992td, Crispino:2007eb}. A closely related discussion about
whether or not uniformly accelerated charges emit radiation  according to inertial observers can be found in Refs.~\cite{Fult}, where the problem is addressed in a classical context.} as a ``theoretical check" of the Fulling-Davies-Unruh (FDU) effect~\cite{Unruh:1976db}, whose implications in Quantum Field Theory (QFT) are still matter of study~\cite{Becattini:2017ljh, Goto:2018ijz}. For technical reasons, the analysis of Refs.~\cite{Matsas:1999jx,Matsas:1999jx3} was performed in two dimensions and taking the neutrino to be massless. Subsequently, Suzuki and Yamada~\cite{Suzuki:2002xg} confirmed these results by extending the discussion to the four dimensional case with massive neutrino. 

Recently, Ahluwalia, Labun and Torrieri~\cite{Aluw} made the very intriguing observation that neutrino mixing can have non-trivial consequences in this context. They indeed found that the decay rates in the two frames could possibly not agree due to mixing terms: in particular, this  happens when  neutrino mass eigenstates are taken as asymptotic states in the comoving frame, a choice which is compatible with the Kubo-Martin-Schwinger (KMS) condition for thermal states~\cite{Haag:1967sg, Crispino:2007eb}. On the other hand, the authors of Ref.~\cite{Aluw} also affirm that the choice of flavor states in the above calculation would lead to an equality of the two decay rates, but in that case the accelerated neutrino vacuum would not be thermal, contradicting the essential characteristic of the FDU effect.  They finally conclude  that such a contradiction has to be solved experimentally.

Motivated by the idea that the above question must instead be settled at a theoretical level in order to guarantee the consistency of the theory in conformity with the principle of General Covariance, in this paper we carefully analyze the proton decay process in the presence of neutrino mixing. We show that in Ref.~\cite{Aluw} calculations performed in the laboratory frame neglected an important contribution which here is explicitly evaluated. Then, we prove that the choice of neutrinos with definite masses as asymptotic states (in the comoving frame) inevitably leads to a disagreement of the two decay rates. Finally, we consider the case  when flavor states are taken into account: here technical difficulties arise which do not allow for an exact evaluation of the decay rates. However, adopting an appropriate approximation, we show that they perfectly match again. 

These conclusions are in line with results on the quantization for mixed neutrino fields, whereby flavor states are rigorously defined as eigenstates of the leptonic charge operators~\cite{Blasone:2001qa}. Although the usual Pontecorvo states turn out to be a good approximation of the exact flavor eigenstates in the ultrarelativistic limit, the Hilbert space associated to flavor neutrinos is actually orthogonal to the one for massive neutrinos~\cite{Blasone:1995zc}. Consistency with the Standard Model (SM) requires conservation (at tree level) of leptonic number in the charged current weak interaction vertices, thus ruling out the choice of neutrino mass eigenstates as asymptotic states. 

Results of the present paper corroborate this view, although further investigation is needed to go beyond the aforementioned approximation. In particular, when employing the exact neutrino flavor states, one should take into account the non-thermal character of Unruh radiation, as recently discussed in Refs.~\cite{Blasone:2017nbf,Blasone:2018byx}.

The paper is organized as follows. Section II is devoted to briefly review the standard calculation of the proton decay rate both in the inertial and comoving frame. For this purpose, we closely follow Ref.~\cite{Suzuki:2002xg}. In Section III we analyze the same process in the context  of neutrino flavor mixing. Working within a suitable framework, we show the decay rates to agree with each other. Our results shall thus be critically compared with the ones of Ref.~\cite{Aluw}, where a contradiction is instead highlighted. Section IV contains conclusions and an outlook at future developments of present work.

Throughout the paper, we use natural units $\hslash=c=1$  and the Minkowski metric with the conventional timelike signature:
\begin{equation}
\eta_{\mu\nu}\, =\, {\rm diag}(+1,-1,-1,-1).
\label{eqn:flatmetr}
\end{equation}

\section{Decay of accelerated protons: a brief review}
\label{sect1}
In this Section we discuss the decay of accelerated protons both in the laboratory and comoving frame. Throughout the whole analysis, neutron $|n\rangle$ and proton $|p\rangle$ are considered as excited and unexcited states of the nucleon, respectively. Moreover, we assume that they are energetic enough to have a well defined trajectory. As a consequence,  the current-current interaction of Fermi theory can be treated with a classical hadronic current $\hat{J}^{\mu}_{\ell}\hat{J}_{h,\mu}\,\rightarrow\,\hat{J}^{\mu}_{\ell}\hat{J}^{(cl)}_{h,\mu}$, where 
\be
\label{Jcl}
\hat{J}^{(cl)}_{h,\mu}\ =\ \hat{q}(\tau)\hspace{0.2mm}u_{\mu}\hspace{0.2mm}\delta(x)\hspace{0.2mm}\delta(y)\hspace{0.2mm}\delta(u-a^{-1})\,.
\ee
Here $u\,=\,a^{-1}\,=\,\mathrm{const}$ is the spatial Rindler coordinate describing the world line of the uniformly accelerated nucleon with proper acceleration $a$, and $\tau\,=\,v/a$ is its proper time, with $v$ being the Rindler time coordinate. The nucleon's four-velocity $u^{\mu}$ is given by
\be
u^{\mu}\ =\ (a, 0, 0, 0), \qquad u^{\mu}\ =\ (\sqrt{a^2t^2+1}, 0, 0, a\hspace{0.2mm}t)\,,
\ee  
in Rindler and Minkowski coordinates, respectively\footnote{We assume that the proton is accelerated along the $z$-direction. Hence, the Rindler coordinates $(v, x, y, u)$  are related with the Minkowski coordinates $(t, x, y, z)$ by: $t=u\sinh{v}$, $z=u\cosh{v}$, with $x$ and $y$ left unchanged.}. 
According to Refs.~\cite{Matsas:1999jx, Birrell}, the Hermitian monopole $\hat{q}(\tau)$  is defined as
\be
\hat{q}(\tau)\ \equiv\ e^{i\hat{H}\tau}\hspace{0.2mm}\hat {q}_0\hspace{0.2mm}e^{-i\hat{H}\tau},
\ee 
where $\hat{H}$ is the nucleon Hamiltonian and $\hat {q}_0$ is related to the Fermi constant $G_F$ by
\be
G_F\ \equiv\ \langle p\hspace{0.2mm}|\hspace{0.2mm}\hat q_0\hspace{0.2mm}|\hspace{0.2mm}n\rangle.
\ee

Next, the minimal coupling of the electron $\hat{\Psi}_{e}$ and neutrino $\hat\Psi_{\nu_e}$ fields to the nucleon current $\hat{J}^{(cl)}_{h,\mu}$ can be expressed through the Fermi action
\be
\label{eqn:Fermiaction}
\hat{S}_{I}\ =\ \hspace{-0.5mm}\int d^{4}x\hspace{0.2mm}\sqrt{-g}\hspace{0.3mm}\hat{J}^{(cl)}_{h,\mu}\left(\hat{\overline{\Psi}}_{\nu_e}\gamma^{\mu}\hat{\Psi}_{e}\, +\, \hat{\overline{\Psi}}_{e}\gamma^{\mu}\hat{\Psi}_{\nu_e}\right)\hspace{-0.2mm},
\ee
where $g\equiv \mathrm{det}(g_{\mu\nu})$ and $\gamma^\mu$ are the gamma matrices in Dirac representation (see, e.g., Ref.~\cite{Itzykson}).

\subsection{Inertial frame calculation}
\label{subsect1}
Let us firstly analyze the decay process in the inertial frame. In this case, the proton is accelerated by an external field and converts into a neutron by emitting a positron and a neutrino, according to 
\be
p\,\rightarrow\, n\, +\, e^{+}\,+\,\nu_e\,.
\ee
In order to calculate the transition rate, we quantize fermionic fields in the usual way \cite{Matsas:1999jx2}:
\be
\label{inertexp}
\hat{\Psi}(t,\textbf{x})\ =\ \sum_{\sigma=\pm}\int d^3k\left[\hat{b}_{\textbf{k}\sigma}\psi_{\bf{k}\sigma}^{(+\omega)}(t,\textbf{x})\,+\,\hat{d}_{\bf{k}\sigma}^{\dagger}\psi_{-\textbf{k}-\sigma}^{(-\omega)}(t,\textbf{x})\right],
\ee
where $\textbf{x}\,\equiv\,(x,y,z)$. Here we have denoted by $\hat{b}_{\textbf{k}\sigma}$ ($\hat{d}_{\textbf{k}\sigma}$) the canonical annihilation operators of fermions (antifermions) with momentum $\textbf{k}\,\equiv\,(k^x, k^y, k^z)$, polarization $\sigma\,=\,\pm$ and frequency $\omega\,=\,\sqrt{\textbf{k}^{2}+m^{2}}\,>\,0$, $m$ being the mass of the field.  The modes $\psi_{\textbf{k}\sigma}^{(\pm\omega)}$ are positive and negative energy solutions of the Dirac equation in Minkowski spacetime:
\be
\big(i\gamma^{\mu}\partial_{\mu}\,-\,m\big)\psi_{\textbf{k}\sigma}^{(\pm\omega)}(t, \textbf{x})\ =\ 0.
\ee
In the adopted representation of $\gamma$ matrices, they take the form~\cite{Matsas:1999jx2}
\be
\label{modes}
\psi_{\textbf{k}\sigma}^{(\pm\omega)}(t, \textbf{x}) \ =\ \frac{e^{i(\mp\omega t\,+\,\textbf{k}\cdot\textbf{x})}}{2^2\hspace{0.2mm}\pi^{\frac{3}{2}}}\hspace{0.2mm}\hspace{0.2mm}u_{\sigma}^{(\pm\omega)}(\textbf{k}),
\ee 
where
\be
u_{+}^{(\pm\omega)}(\textbf{k})\ =\ \frac{1}{\sqrt{\omega(\omega\pm m)}}
\begin{pmatrix}
m\,\pm\,\omega \\
0 \\
k^z \\
k^x\,+\,ik^y
\end{pmatrix},\qquad
u_{-}^{(\pm\omega)}(\textbf{k})\ =\ \frac{1}{\sqrt{\omega(\omega\pm m)}}
\begin{pmatrix}
0 \\
m\,\pm\,\omega \\
k^x\,-\,ik^y \\
-\hspace{0.2mm}k^z
\end{pmatrix}.
\ee
It is easy to show that the modes Eq.~(\ref{modes}) are orthonormal with respect to the inner product 
\be
\label{innprod}
\left\langle\psi_{\textbf{k}\sigma}^{(\pm\omega)},\psi_{\textbf{k}'\sigma'}^{(\pm\omega')}\right\rangle=\int_{\Sigma}d\Sigma_{\mu}\,\overline{\psi}_{\textbf{k}\sigma}^{(\pm\omega)}\gamma^{\mu}\psi_{\textbf{k}'\sigma'}^{(\pm\omega')}\ =\ \delta_{\sigma\sigma'}\delta^3(\textbf{k}-\textbf{k}')\delta_{\pm\omega\pm\omega'},
\ee
where $\overline{\psi}\,=\,\psi^{\dagger}\gamma^{0}$, $d\Sigma_{\mu}\,=\,n_{\mu}d\Sigma$, with $n_{\mu}$ being a unit vector orthogonal to the arbitrary spacelike hypersurface $\Sigma$ and pointing to the future. 

Next, by using the definition Eq.~(\ref{eqn:Fermiaction}) of the Fermi action and expanding leptonic fields according to Eq.~(\ref{inertexp}), we obtain the following expression for the tree-level transition amplitude:
\begin{equation}
\mathcal{A}^{p\rightarrow n}_{in}\ \equiv\ \langle n|\otimes\langle e_{k_{e}\sigma_{e}}^{+},\nu_{k_{\nu}\sigma_{\nu}}|\hat{S}_{I}|0\rangle\otimes|p\rangle\ =\ \frac{G_F}{2^4\pi^3}\,\mathcal{I}_{\sigma_\nu\sigma_e}(\omega_\nu, \omega_e),
\label{tramp}
\end{equation}
where 
\be
\label{I}
\mathcal{I}_{\sigma_\nu\sigma_e}(\omega_\nu, \omega_e)\ =\ \hspace{-0.5mm}\int_{-\infty}^{+\infty}\hspace{-2.3mm}d\tau\, e^{i\big[\Delta m\hspace{0.2mm}\tau\,+\,a^{-1}\left(\omega_\nu\,+\,\omega_{e}\right)\sinh a\tau\,-\,a^{-1}\left(k_\nu^z\,+\,k^z_{e}\right)\cosh a\tau\big]}u_{\mu}\left[\bar{u}_{\sigma_\nu}^{(+\omega_\nu)}\gamma^{\mu}{u}_{-\sigma_e}^{(-\omega_e)}\right].
\ee
Here $\Delta m$ is the difference between the nucleon masses. By defining the differential transition rate as
\begin{eqnarray}
\label{dtr}
\nonumber
\frac{d^{6}\mathcal{P}_{in}^{p\rightarrow n}}{d^3k_{\nu}\,d^3k_{e}}\,\equiv\,\sum_{\sigma_{\nu},\sigma_e}\left|\mathcal{A}_{in}^{p\rightarrow n}\right|^{2} &\,=\,&\frac{G_F^2}{2^8\pi^6}\int_{-\infty}^{+\infty}\hspace{-2mm}d\tau_1\int_{-\infty}^{+\infty}\hspace{-2mm}d\tau_2\hspace{0.3mm}u_{\mu}u_{\nu}\sum_{\sigma_\nu,\sigma_e}\left[\bar{u}_{\sigma_\nu}^{(+\omega_\nu)}\gamma^{\mu}{u}_{-\sigma_e}^{(-\omega_e)}\right]\left[\bar{u}_{\sigma_\nu}^{(+\omega_\nu)}\gamma^{\nu}{u}_{-\sigma_e}^{(-\omega_e)}\right]^*\\[2mm]
&&\times \,e^{i\big[\Delta m(\tau_1-\tau_2)+a^{-1}\left(\omega_\nu+\omega_{e}\right)(\sinh a\tau_1-\sinh a\tau_2)-a^{-1}\left(k_\nu^z+k^z_{e}\right)(\cosh a\tau_1-\cosh a\tau_2)\big]}\,,
\end{eqnarray}
the total transition rate is simply given by
\be
\label{ttr}
\Gamma_{in}^{p\rightarrow n}\ =\ \mathcal{P}_{in}^{p\rightarrow n}/T\,,
\ee 
where $T\,=\,\int_{-\infty}^{+\infty}ds$ is the nucleon proper time. The above integrals can be  solved by introducing the new variables 
\be
\label{newvariab}
\tau_1\ =\ s\,+\,\xi/2,\qquad \tau_2\ =\ s\,-\,\xi/2\,,
\ee
and using the spin sum
\begin{eqnarray}
\label{usumspin}
\nonumber
&&u_{\mu}u_{\nu}\sum_{\sigma_\nu,\sigma_e}\left[\bar{u}_{\sigma_\nu}^{(+\omega_\nu)}\gamma^{\mu}{u}_{-\sigma_e}^{(-\omega_e)}\right]\left[\bar{u}_{\sigma_\nu}^{(+\omega_\nu)}\gamma^{\nu}{u}_{-\sigma_e}^{(-\omega_e)}\right]^*\\[2mm]
&&=\frac{2^2}{\omega_{\nu}\omega_e}\Big[\big(\omega_\nu\omega_e\,+\,k^z_\nu k^z_e\big)\cosh 2as\,-\,\big(\omega_\nu k^z_e\,+\,\omega_ek^z_\nu\big)\sinh 2as\,+\,\big(k^x_\nu k^x_e\,+\,k^y_\nu k^y_e\,-\,m_\nu m_e\big)\cosh a\xi\Big].
\end{eqnarray}
By explicit calculation, we obtain
\be\label{gammain}
\Gamma_{in}^{p\rightarrow n}\ =\ \frac{G_F^2}{a\,\pi^6e^{\pi\Delta m/a}}\int d^3 k_\nu\int d^3 k_e \left[K_{2i\Delta m/a}\left(\frac{2(\omega_\nu\,+\,\omega_e)}{a}\right)\,+\,\frac{m_\nu m_e}{\omega_\nu\omega_e}\,\mathrm{Re}\left\{K_{2i\Delta m/a+2}\left(\frac{2(\omega_\nu\,+\,\omega_e)}{a}\right)\right\}\right].
\ee
The analytic evaluation of the integral Eq.~(\ref{gammain}) can be found in Ref.~\cite{Suzuki:2002xg}.

\subsection{Comoving frame calculation}
We now analyze the same decay process in the proton comoving frame. As well-known, the natural manifold to describe phenomena for uniformly accelerated observers is the Rindler wedge, i.e., the Minkowski spacetime region defined by $z>\left|t\right|$. Within such a manifold, fermionic fields are expanded in terms of the positive and negative frequency solutions of the Dirac equation with respect to the boost Killing vector $\partial/\partial v$~\cite{Suzuki:2002xg}:
\be 
\label{Rindexp}
\hat{\Psi}(v,\textbf{x})\ =\ \sum_{\sigma=\pm}\int_{0}^{+\infty}\hspace{-2mm}d\omega\int d^2k\left[\hat{b}_{\textbf{w}\sigma}\psi^{(+\omega)}_{\textbf{w}\sigma}(v,\textbf{x})\,+\,\hat{d}_{\textbf{w}\sigma}^{\dagger}\psi^{(-\omega)}_{\textbf{w}-\sigma}(v,\textbf{x})\right],
\ee
where now $\textbf{x}\,\equiv\,(x,y,u)$ and $\textbf{w}\,\equiv\,(\omega, k^x, k^y)$. We recall that the Rindler frequency $\omega$ may assume arbitrary positive real values. In particular, unlike the inertial case, there are massive Rindler particles with zero frequency. 

The modes $\psi_{\textbf{k}\sigma}^{(\pm\omega)}$ in Eq.~(\ref{Rindexp}) are positive and negative energy solutions of the Dirac equation in Rindler spacetime:
\be
\label{DR}
(i\gamma^{\mu}_R\widetilde\nabla_\mu\,-\,m)\hspace{0.2mm}\psi_{\textbf{w}\sigma}^{(\omega)}(v, \textbf{x})\ =\ 0,
\ee
where
\be
\begin{split}
&\gamma^{\mu}_R\ \equiv\ (e_\nu)^{\mu}\gamma^{\nu},\qquad (e_0)^{\mu}\ =\ u^{-1}\delta_0^{\mu},\qquad (e_i)^{\mu}\ =\ \delta_i^{\mu},\\[2mm]
&\tilde\nabla_{\mu}\ \equiv\ \partial_{\mu}\,+\,\frac{1}{8}\left[\gamma^{\alpha}, \gamma^\beta\right]\hspace{-0.3mm}{\left(e_\alpha\right)}^\lambda\nabla_\mu(e_\beta)_\lambda.
\label{DiracRind}
\end{split}
\ee
By virtue of these relations and using the Rindler coordinates, Eq.~(\ref{DR}) becomes
\be
i\frac{\partial \psi^{(\omega)}_{\textbf{w}\sigma}(v, \textbf{x})}{\partial v}\ =\ \left(\gamma^0mu\,-\,\frac{i\alpha^3}{2}\,-\,iu\alpha^i\partial_i\right)\hspace{-0.2mm}\psi^{(\omega)}_{\textbf{w}\sigma}(v, \textbf{x}),\qquad \alpha^i\,=\,\gamma^0\gamma^i, \,\,\,i=1,2,3
\ee
whose solutions can be written in the form~\cite{Suzuki:2002xg}
\be
\label{modesRind}
\psi_{\textbf{w}\sigma}^{(\omega)}(v, \textbf{x})\ =\ \frac{e^{i(-\omega v/a\,+\,k_\alpha x^\alpha)}}{{(2\pi)}^{\frac{3}{2}}}\hspace{0.2mm}u_{\sigma}^{(\omega)}\hspace{0.2mm}(u,\textbf{w}),\qquad \alpha=1,2,
\ee 
with
\be
u_{+}^{(\omega)}(u,\textbf{w})=
N\begin{pmatrix}
i\hspace{0.2mm}l\hspace{0.1mm}K_{i\omega/a-1/2}(u\hspace{0.2mm}l)\,+\,m\hspace{0.2mm}K_{i\omega/a+1/2}(ul) \\[3mm]
-(k^x\,+\,ik^y)K_{i\omega/a+1/2}(ul) \\[3mm]
i\hspace{0.2mm}l\hspace{0.1mm}K_{i\omega/a-1/2}(u\hspace{0.2mm}l)\,-\,m\hspace{0.2mm}K_{i\omega/a\,+\,1/2}(ul) \\[3mm]
-(k^x+ik^y)K_{i\omega/a+1/2}(ul)
\end{pmatrix}\hspace{-0.6mm},
\quad\,
u_{-}^{(\omega)}(u,\textbf{w})=
N\begin{pmatrix}
(k^x\,-\,ik^y) K_{i\omega+1/2}(ul) \\[3mm]
i\hspace{0.2mm}l\hspace{0.1mm}K_{i\omega/a-1/2}(u\hspace{0.2mm}l)\,+\,m\hspace{0.2mm}K_{i\omega/a+1/2}(ul) \\[3mm]
-(k^x\,-\,ik^y) K_{i\omega+1/2}(ul) \\[3mm]
-i\hspace{0.2mm}l\hspace{0.1mm}K_{i\omega/a-1/2}(u\hspace{0.2mm}l)\,+\,m\hspace{0.2mm}K_{i\omega/a+1/2}(ul) 
\end{pmatrix}.
\ee
Here we have denoted by $K_{i\omega/a+1/2}(ul)$ the modified Bessel function of the second kind with complex order,  $N\,=\,\sqrt{\frac{a\cosh(\pi\omega/a)}{\pi l}}$ and $l\,=\,\sqrt{m^2+(k^x)^2+(k^y)^2}$. Again, one can verify that the modes in Eq.~(\ref{modesRind}) are normalized with respect to the inner product Eq.~(\ref{innprod}) expressed in Rindler coordinates.
\medskip

As it will be shown, in the comoving frame the proton decay is represented as the combination of the three following processes in terms of the Rindler particles~\cite{Matsas:1999jx}:
\be
\label{threeprocesses}
(i)\quad p^{+}\,+\,e^{-}\,\rightarrow\, n\,+\,\nu_e, \qquad (ii)\quad p^{+}\,+\,\overline{\nu}_e\,\rightarrow\, n\,+\,e^{+}, \qquad (iii)\quad p^{+}\,+\,e^{-}\,+\,\overline{\nu}_e\,\rightarrow\, n.
\ee   
These processes are characterized by the conversion of
protons in neutrons due to the absorption of $e^-$ and $\bar{\nu}_e$, and
emission of $e^+$ and $\bar{\nu}_e$ from and to the FDU thermal bath~\cite{Unruh:1976db}. Since the strategy for calculating the transition amplitude is the same for each of these processes, by way of illustration we shall focus on the first. 

By exploiting the Rindler expansion Eq.~(\ref{Rindexp}) for the electron and neutrino fields, it can be shown that
\begin{equation}
\label{first}
\mathcal{A}^{p\rightarrow n}_{(i)}\ \equiv\ \left\langle n\right|\otimes\langle\nu_{\omega_\nu\hspace{0.2mm}\sigma_{\nu}}|\hspace{0.2mm}\hat{S}_{I}\hspace{0.2mm}|e^{-}_{\omega_{e^-}\hspace{0.2mm}\sigma_{e^-}}\rangle\otimes\left|p\right\rangle\ =\ \frac{G_F}{(2\pi)^2}\,\mathcal{J}_{\sigma_\nu\sigma_e}(\omega_\nu, \omega_e),
\end{equation}
where $\hat{S}_I$ is given by Eq.~(\ref{eqn:Fermiaction}) with $\gamma^\mu$ replaced by the Rindler gamma matrices $\gamma^{\mu}_R$ defined in Eq.~(\ref{DiracRind}) and
\be
\label{eqn:j}
\mathcal{J}_{\sigma_\nu\sigma_e}(\omega_\nu, \omega_e)\ =\ \delta\big(\omega_e-\omega_{\nu}-\Delta m\big)\,\bar{u}_{\sigma_{\nu}}^{(\omega_{\nu})}\gamma^0 u_{\sigma_{e}}^{(\omega_{e})}.
\ee

Now, bearing in mind that the probability for the proton to absorb (emit) a particle of frequency $\omega$ from (to) the thermal bath is $n_{F}(\omega)\,=\,\frac{1}{e^{2\pi\omega/a}\,+\,1}$ $\big(1\,-\,n_{F}(\omega)\big)$~\cite{Matsas:1999jx,Matsas:1999jx2,Matsas:1999jx3},  
the differential transition rate per unit time for the process $(i)$ can be readily evaluated, thus leading to 
\begin{eqnarray}
\label{eqn:diftransraterind}
\frac{1}{T}\frac{d^{6}\mathcal{P}^{p\rightarrow n}_{(i)}}{d\omega_{\nu}\hspace{0.2mm}d\omega_{e}\hspace{0.2mm}d^2k_{\nu}\hspace{0.2mm}d^2k_e}&\equiv&\frac{1}{T}\sum_{\sigma_{\nu},\sigma_e}\big|\mathcal{A}^{p\rightarrow n}_{(i)}\big|^{2}n_{F}(\omega_{e})\big[1\,-\,n_{F}(\omega_\nu)\big]\\[2mm]
\nonumber
&=&\frac{\hspace{0.2mm}G_{F}^{2}}{2^7\pi^{5}}\hspace{0.2mm}\frac{\sum_{\sigma_{\nu},\sigma_e}\big|\bar{u}_{\sigma_{\nu}}^{(\omega_{\nu})}\gamma^0 u_{\sigma_{e}}^{(\omega_{e})}\big|^2\hspace{0.2mm}\delta\left(\omega_e-\omega_{\nu}-\Delta m\right)}{e^{\pi\Delta m/a}\cosh(\pi\omega_{\nu}/a)\cosh(\pi\omega_{e}/a)},
\end{eqnarray}
where $T\,=\,2\pi\delta(0)$ is the total proper time of the proton. In order to finalize the evaluation of the transition rate, we observe that
\bea
\label{spinsum2}
\sum_{\sigma_{\nu},\sigma_e}\big|\bar{u}_{\sigma_{\nu}}^{(\omega_{\nu})}\gamma^0 u_{\sigma_{e}}^{(\omega_{e})}\big|^2&=&\frac{2^4}{(a\,\pi)^2}\cosh(\pi\omega_{\nu}/a)\cosh(\pi\omega_{e}/a)\left[l_{\nu}l_e\Bigl|K_{i\omega_{\nu}/a+1/2}\left(\frac{l_{\nu}}{a}\right)K_{i\omega_{e}/a+1/2}\left(\frac{l_{e}}{a}\right)\Bigr|^2\right.
\\[2mm]
\nonumber
&&\left.+\left(k^x_\nu k^x_e\,+\,k^y_\nu k^y_e\,+\,m_{\nu}m_e\right)\mathrm{Re}\left\{K^2_{i\omega_{\nu}/a-1/2}\left(\frac{l_{\nu}}{a}\right)K^2_{i\omega_{e}/a+1/2}\left(\frac{l_{e}}{a}\right)\right\}\right].
\eea
Using this equation, the differential transition rate for the process $(i)$ takes the form
\bea\nonumber
\label{difftransrate}
\frac{1}{T}\frac{d^{6}\mathcal{P}^{p\rightarrow n}_{(i)}}{d\omega_{\nu}d\omega_{e}d^2k_{\nu}d^2k_e}&\equiv&\frac{G_F^2}{2^3\,a^2\,\pi^7\,e^{\pi\Delta m/a}}\hspace{0.1mm}\delta\left(\omega_e-\omega_{\nu}-\Delta m\right)\left[l_{\nu}l_e\Bigl|K_{i\omega_{\nu}/a+1/2}\left(\frac{l_{\nu}}{a}\right)K_{i\omega_{e}/a+1/2}\left(\frac{l_{e}}{a}\right)\Bigr|^2\right.\\[2mm]
&&\left.+\,m_{\nu}m_e\mathrm{Re}\left\{K^2_{i\omega_{\nu}/a-1/2}\left(\frac{l_{\nu}}{a}\right)K^2_{i\omega_{e}/a+1/2}\left(\frac{l_{e}}{a}\right)\right\}\right].
\eea

Next, by performing similar calculation for the processes $(ii)$ and $(iii)$ and adding up the three contributions, we end up with the following integral expression for the total decay rate in the comoving frame:
\be\label{gammaacc}
\Gamma_{acc}^{p\rightarrow n}\ \equiv\ \Gamma_{(i)}^{p\rightarrow n}\,+\,\Gamma_{(ii)}^{p\rightarrow n}\,+\,\Gamma_{(iii)}^{p\rightarrow n}\ =\ \frac{2\hspace{0.2mm}G_{F}^{2}}{a^2\hspace{0.2mm}\pi^7\hspace{0.2mm}e^{\pi\Delta m/a}}\int_{-\infty}^{+\infty}\hspace{-2mm}d\omega\,\mathcal{R}(\omega),
\ee 
where
\bea\nonumber
\label{R}
\mathcal{R}(\omega)&=&\int d^2k_{\nu}\,l_{\nu}\Bigl|K_{i(\omega-\Delta m)/a+1/2}\left(\frac{l_{\nu}}{a}\right)\Bigr|^2\int d^2k_{e}\,l_{e}\Bigl|K_{i\omega/a+1/2}\left(\frac{l_{e}}{a}\right)\Bigr|^2\\[2mm]
&&+\,m_{\nu}m_e\mathrm{Re}\hspace{0.2mm}\left\{\int d^2k_{\nu}K^2_{i(\omega-\Delta m)/a-1/2}\left(\frac{l_{\nu}}{a}\right)\int d^2k_eK^2_{i\omega/a+1/2}\left(\frac{l_{e}}{a}\right)\right\}.
\eea
The analytic resolution of the integral Eq.~(\ref{gammaacc}) is performed in Ref.~\cite{Suzuki:2002xg}. Comparing this result to the one in the inertial frame (Eq.~(\ref{gammain})), it is possible to show that the resulting expressions for the decay rates perfectly agree with each other, thus corroborating the necessity of the FDU effect for the consistency of QFT.

\section{Proton decay involving mixed neutrinos}
\label{prdec}
So far, in the evaluation of the transition amplitude, we have treated the electron neutrino as a particle with definite mass $m_\nu$. However, it is well-known that neutrinos exhibit flavor mixing: in a simplified two-flavor model, by denoting with $\theta$ the mixing angle, the transformations relating the flavor eigenstates $|\nu_{\ell}\rangle$ ($\ell\,=\,e, \mu$) and mass eigenstates $|\nu_i\rangle$ ($i\ =\ 1, 2$) are determined by the Pontecorvo unitary mixing matrix\footnote{Note that the number of neutrino generations does not affect the results of our analysis.}~\cite{Bilenky:1978nj}

\be
\begin{pmatrix}
\label{Pontec}
|\nu_e\rangle\\
|\nu_\mu\rangle
\end{pmatrix}\,=\, \begin{pmatrix}
\cos\theta&\sin\theta\\
-\sin\theta&\cos\theta
\end{pmatrix}
\begin{pmatrix}
|\nu_1\rangle\\
|\nu_2\rangle
\end{pmatrix}.
\ee

Along the line of Ref.~\cite{Aluw}, the question thus arises whether such a transformation is consistent with the framework of Sec.~\ref{sect1}.

\subsection{Inertial frame}
\label{inerframemix}
Let us then implement the Pontecorvo rotation Eq.~(\ref{Pontec}) on both the neutrino fields and states appearing in Eq.~(\ref{tramp}).  Note that in Ref.~\cite{Aluw} this step is missing in the inertial frame calculation since $\hat\Psi_{\nu_e}$ is treated as a free-field even when taking into account flavor mixing, and indeed the same result as in the case of unmixed fields is obtained.  We  explicitly show that the decay rate exhibits a dependence on  $\theta$ in the inertial frame, a feature which is not present in the analysis of Ref.~\cite{Aluw}. 

By assuming equal momenta and polarizations for the two neutrino mass eigenstates, the transition amplitude Eq.~(\ref{tramp}) now becomes
\begin{equation}
\mathcal{A}_{in}^{p\rightarrow n}\ =\ \frac{G_F}{2^4\pi^3}\Big[\cos^2\theta\, \mathcal{I}_{\sigma_\nu\sigma_e}(\omega_{\nu_1},\omega_e)\,+\, \sin^2\theta\, \mathcal{I}_{\sigma_\nu\sigma_e}(\omega_{\nu_2},\omega_e)\Big],
\end{equation}
where $\mathcal{I}_{\sigma_\nu\sigma_e}(\omega_{\nu_j},\omega_e)$, $j\,=\,1,2$, is defined as in Eq.~(\ref{I}) for each of the two mass eigenstates, and we have rotated the electron neutrino field according to 
\be
\label{rotfield}
\hat\Psi_{\nu_e}(t,\textbf{x})\ =\ \cos\theta\hspace{0.4mm}\hat\Psi_{\nu_1}(t,\textbf{x})\,+\,\sin\theta\hspace{0.4mm}\hat\Psi_{\nu_2}(t,\textbf{x}).
\ee 
Using Eq.~(\ref{dtr}), the differential transition rate takes the form
\begin{eqnarray}\nonumber
\frac{d^6\mathcal{P}_{in}^{p\rightarrow n}}{d^3k_\nu\hspace{0.2mm} d^3k_e}&=&\sum_{\sigma_\nu,\sigma_e}\frac{{G_F}^2}{2^8\pi^6}\hspace{-0.5mm}\left\{\cos^4\theta\,{\big|\mathcal{I}_{\sigma_\nu\sigma_e}(\omega_{\nu_1},\omega_e)\big|}^2\,+\,\sin^4\theta\,{\big|\mathcal{I}_{\sigma_\nu\sigma_e}(\omega_{\nu_2},\omega_e)\big|}^2\right.\\[2mm]
&&+\,\cos^2\theta\sin^2\theta\Big[\mathcal{I}_{\sigma_\nu\sigma_e}(\omega_{\nu_1},\omega_e)\,\mathcal{I}^{\hspace{0.2mm}*}_{\sigma_\nu\sigma_e}(\omega_{\nu_2},\omega_e)\,+\,\mathrm{c.c.}\Big]\Big\}\,.
\end{eqnarray}
The total decay rate $\Gamma_{in}^{p\rightarrow n}$  is obtained after inserting this equation  into the definition Eq.~(\ref{ttr}):
\be
\label{eqn:inertresultat}
\Gamma^{p\rightarrow n}_{in}\ =\ \cos^4\theta\, \Gamma^{p\rightarrow n}_{1}\,+\,\sin^4\theta\,\Gamma^{p\rightarrow n}_{2}\,+\,\cos^2\theta\sin^2\theta\,\Gamma^{p\rightarrow n}_{12},
\ee
where we have introduced the shorthand notation
\begin{equation}
\label{integral}
\Gamma^{p\rightarrow n}_{j}\ \equiv\ \frac{1}{T}\sum_{\sigma_\nu,\sigma_e}\frac{{G_F}^2}{2^8\pi^6}\int d^3k_\nu\int d^3k_e\,{\big|\mathcal{I}_{\sigma_\nu\sigma_e}(\omega_{\nu_j},\omega_e)\big|}^2,\qquad j=1,2,
\end{equation}
and 
\begin{equation}
\label{integral12}
\Gamma^{p\rightarrow n}_{12}\ \equiv\ \frac{1}{T}\sum_{\sigma_\nu,\sigma_e}\frac{{G_F}^2}{2^8\pi^6}\int d^3k_\nu\int d^3k_e\Big[\mathcal{I}_{\sigma_\nu\sigma_e}(\omega_{\nu_1},\omega_e)\,\mathcal{I}^{\hspace{0.2mm}*}_{\sigma_\nu\sigma_e}(\omega_{\nu_2},\omega_e)\,+\,\mathrm{c.c.}\Big].
\end{equation}

We observe that, for $\theta\rightarrow 0$, the obtained result correctly reduces to Eq.~(\ref{gammain}), as it should be in absence of mixing. Unfortunately, due to technical difficulties in the evaluation of the integral Eq.~(\ref{integral12}), at this stage we are not able to give the exact expression  of the inertial decay rate Eq.~(\ref{eqn:inertresultat}). A preliminary result, however, can be obtained in the limit of small neutrino mass difference $\frac{\delta m}{m_{\nu_1}}\,\equiv\,\frac{m_{\nu_2}\,-\,m_{\nu_1}}{m_{\nu_1}}\,\ll\, 1$. In this case, indeed, we can expand  $\Gamma^{p\rightarrow n}_{12}$ according to
\be
\label{approxima}
\Gamma^{p\rightarrow n}_{12}\ =\ 2\Gamma^{p\rightarrow n}_1\,+\,\frac{\delta m}{m_{\nu_1}}\,\Gamma^{(1)}\,+\, \mathcal{O}\left(\frac{\delta m^2}{m^2_{\nu_1}}\right),
\ee
where $\Gamma^{p\rightarrow n}_1$ is defined as in Eq.~(\ref{integral}) and we have denoted by $\Gamma^{(1)}$ the first-order term of the Taylor expansion. The explicit expression of $\Gamma^{(1)}$ is rather awkward to exhibit. Nevertheless, for $m_{\nu_1}\rightarrow 0$, it can be  substantially simplified, thus giving
\bea
\label{firstorder}
\nonumber
\frac{\Gamma^{(1)}}{m_{\nu_1}}\,=\,\frac{1}{T}\frac{{G_F}^2\,m_e}{2^7\pi^6}\int \frac{d^3k_\nu}{|k_\nu|}\int \frac{d^3 k_e}{\omega_e} \int_{-\infty}^{+\infty}\hspace{-2mm}ds\int_{-\infty}^{+\infty}\hspace{-2mm}d\xi \cosh a\xi\left[e^{i\big\{\Delta m\,\xi+\frac{2\sinh a\xi/2}{a}\big[\left(|k_\nu|+\omega_{e}\right)\cosh as-\left(k_\nu^z+k^z_{e}\right)\sinh as\big]\big\}}\,+\,\mathrm{c.c.}\right],\\
\eea
where $s$ and $\xi$ are defined in Eq.~(\ref{newvariab}). By performing a boost along the $z$-direction: 
\be
{k'}^{\hspace{0.2mm}x}_{\hspace{0.2mm}\ell}\ =\ {k}^{\hspace{0.2mm}x}_{\hspace{0.2mm}\ell},\qquad {k'}^{\hspace{0.2mm}y}_{\hspace{0.2mm}\ell}\ =\ {k}^{\hspace{0.2mm}y}_{\hspace{0.2mm}\ell},\qquad {k'}^{\hspace{0.2mm}z}_{\hspace{0.2mm}\ell}\, =\,- \hspace{0.01mm}\omega_\ell\sinh as\,+\,{k}^{\hspace{0.2mm}z}_{\hspace{0.2mm}\ell}\cosh as,\qquad\quad \ell\,=\,\nu_1,e,
\ee
Eq.~(\ref{firstorder}) can be cast in the form
\be\label{besselk}
\frac{\Gamma^{(1)}}{m_{\nu_1}}\ =\ \lim_{\varepsilon\rightarrow 0}\hspace{0.3mm}\frac{2\,G_F^2\,m_e}{a\,\pi^6\,e^{\pi\Delta m/a}}\int\frac{d^3k_\nu}{\omega_\varepsilon}\int\frac{d^3k_e}{\omega_e}\mathrm{Re}\left\{K_{2i\Delta m/a+2}\left(\frac{2(\omega_\varepsilon\,+\,\omega_e)}{a}\right)\right\},
\ee
where $\omega_\varepsilon\,=\,\sqrt{\textbf{k}^2_\nu\,+\,\varepsilon^2}$, with $\varepsilon$ acting as a regulator. In order to perform $k$-integration, we use the following representation of the modified Bessel function:
\be
K_\mu(z)\ =\ \frac{1}{2}\int_{C_1}\frac{ds}{2\pi i}\Gamma(-s)\Gamma(-s-\mu)\left(\frac{z}{2}\right)^{2s+\mu},
\ee
where $\Gamma$ is the Euler's Gamma function. $C_1$ is the path in the complex plane including all the poles of $\Gamma(-s)$ and $\Gamma(-s-\mu)$, chosen in such a way that the integration with respect to the momentum variables does not diverge~\cite{Suzuki:2002xg}.

Using spherical coordinates, Eq.~(\ref{besselk}) becomes
\bea\nonumber
\frac{\Gamma^{(1)}}{m_{\nu_1}}&=&\lim_{\varepsilon\rightarrow 0}\hspace{0.3mm}\frac{2^3\,G_F^2\,m_e}{a\,\pi^4\,e^{\pi\Delta m/a}}\int_{0}^{+\infty}\hspace{-2mm}dk_\nu\hspace{0.3mm}\frac{k^2_\nu}{\omega_\varepsilon}\int_{0}^{+\infty}\hspace{-2mm}dk_e\hspace{0.3mm}\frac{k^2_e}{\omega_e}\int_{C_s}\frac{ds}{2\pi i}\hspace{-0.2mm}\left(\frac{\omega_\varepsilon\,+\,\omega_e}{a}\right)^{2s}\\[2mm]
&&\times\left[\Gamma\left(-s\hspace{0.2mm}+\hspace{0.2mm}\frac{i\Delta m}{a}\hspace{0.2mm}+\hspace{0.2mm}1\right)\Gamma\left(-s\hspace{0.2mm}-\hspace{0.2mm}\frac{i\Delta m}{a}\hspace{0.2mm}-\hspace{0.2mm}1\right)\,+\,\Gamma\left(-s\hspace{0.2mm}+\hspace{0.2mm}\frac{i\Delta m}{a}\hspace{0.2mm}-\hspace{0.2mm}1\right)\Gamma\left(-s\hspace{0.2mm}-\hspace{0.2mm}\frac{i\Delta m}{a}\hspace{0.2mm}+\hspace{0.2mm}1\right)\right].
\eea
Let us observe at this point that~\cite{Suzuki:2002xg}
\be
\left(\frac{\omega_\varepsilon\,+\,\omega_e}{a}\right)^{2s}\, =\, \int_{C_2}\frac{dt}{2\pi i}\frac{\Gamma(-t)\Gamma(t\hspace{0.2mm}-\hspace{0.2mm}2s)}{\Gamma(-2s)}\left(\frac{\omega_\varepsilon}{a}\right)^{-t+2s}\left(\frac{\omega_e}{a}\right)^t,
\ee
where $C_2$ is the contour in the complex plane separating the poles of $\Gamma(-t)$ from the ones of $\Gamma(t-2s)$. Exploiting this relation and properly redefining the integration variables, we finally obtain
\bea
\label{lastin}
\frac{\Gamma^{(1)}}{m_{\nu_1}}&=&\lim_{\varepsilon\rightarrow 0}\hspace{0.3mm}\frac{G_F^2\,m_e\,a^3}{\pi^3\,e^{\pi\Delta m/a}}\int_{C_s}\frac{ds}{2\pi i}\int_{C_t}\frac{dt}{2\pi i}\left(\frac{\varepsilon}{a}\right)^{2s+2}\left(\frac{m_e}{a}\right)^{2t+2}\frac{\Gamma(-2s)\Gamma(-2t)\Gamma(-t\,-\,1)\Gamma(-s\,-\,1)}{\Gamma(-s\,+\,\frac{1}{2})\Gamma(-t\,+\,\frac{1}{2})\Gamma(-2s\,-\,2t)}\\[2mm]\nonumber
&&\times\left[\Gamma\left(-s\,-\,t\,+\,1\,+\,i\hspace{0.2mm}\frac{\Delta m}{a}\right)\Gamma\left(-s\hspace{0.2mm}-\hspace{0.2mm}t\hspace{0.2mm}-\hspace{0.2mm}1\hspace{0.2mm}-\hspace{0.2mm}i\hspace{0.2mm}\frac{\Delta m}{a}\right)+\Gamma\left(-s\hspace{0.2mm}-\hspace{0.2mm}t\hspace{0.2mm}+\hspace{0.2mm}1\hspace{0.2mm}-\hspace{0.2mm}i\hspace{0.2mm}\frac{\Delta m}{a}\right)\Gamma\left(-s\hspace{0.2mm}-\hspace{0.2mm}t\hspace{0.2mm}-\hspace{0.2mm}1\hspace{0.2mm}+\hspace{0.2mm}i\hspace{0.2mm}\frac{\Delta m}{a}\right)\right].
\eea
where the contour $C_{s(t)}$ includes all poles of gamma functions in $s$ ($t$) complex plane.

From Eqs.~(\ref{approxima}) and (\ref{lastin}), we thus infer that the off-diagonal term $\Gamma_{12}^{p\rightarrow n}$ is non-vanishing, thereby leading to a structure of the inertial decay rate Eq.~(\ref{eqn:inertresultat}) that is different from the corresponding one in Ref.~\cite{Aluw}.

\subsection{Comoving frame calculation}
\label{sect2}
Let us now extend the above discussion to the proton comoving frame. As done in the inertial case, we require the asymptotic neutrino states to be flavor eigenstates (the choice of mass eigenstates would inevitably lead to a contradiction, as shown in the Appendix). Note that the same assumption is contemplated  also in Ref.~\cite{Aluw}. In spite of this, those authors exclude such an alternative on the basis of the KMS condition, claiming that  the accelerated neutrino vacuum must be a thermal state of neutrinos with definite masses rather than definite flavors. Actually, this argument does not apply, at least  within the first-order approximation we are dealing with (see Eq.~(\ref{approxima})). Indeed, as shown in Refs.~\cite{Blasone:2017nbf, Blasone:2018byx}, non-thermal corrections to the Unruh spectrum for flavor (mixed) neutrinos only appear at orders higher than $\mathcal{O}\left(\frac{\delta m}{m}\right)$. 

Relying on these considerations, let us evaluate the decay rate in the comoving frame. A straightforward calculation leads to the following expression for the transition amplitude Eq.~(\ref{first}):
\be
\mathcal{A}_{(i)}^{p\rightarrow n} \ =\ \frac{G_{F}}{(2\pi)^2}\left[\cos^2\theta\mathcal{J}^{(1)}_{\sigma_\nu\sigma_e}(\omega_\nu, \omega_e)\,+\, \sin^2\theta\mathcal{J}^{(2)}_{\sigma_\nu\sigma_e}(\omega_\nu, \omega_e)\right],
\ee
where $\mathcal{J}^{(j)}_{\sigma_\nu\sigma_e}(\omega_\nu, \omega_e)$, $j\,=\,1,2$, is defined as in Eq.~(\ref{eqn:j}) for each of the two neutrino mass eigenstates. The differential transition rate per unit time thus reads
\begin{eqnarray}
\label{difmixrindright}
\nonumber
\frac{1}{T}\frac{d^{6}\mathcal{P}^{p\rightarrow n}_{(i)}}{d\omega_{\nu}\hspace{0.2mm}d\omega_{e}\hspace{0.2mm}d^2k_{\nu}\hspace{0.2mm}d^2k_e}&=&\frac{1}{T}\frac{\hspace{0.2mm}G_{F}^{2}}{2^6\pi^{4}}\hspace{0.2mm}\frac{1}{e^{\pi\Delta m/a}\cosh(\pi\omega_{\nu}/a)\cosh(\pi\omega_{e}/a)}\sum_{\sigma_{\nu},\sigma_e}\Bigl\{\cos^4\theta\,\big|\mathcal{J}^{(1)}_{\sigma_\nu\sigma_e}(\omega_\nu, \omega_e)\big|^2\\[2mm]
&&+\,\sin^4\theta\,\big|\mathcal{J}^{(2)}_{\sigma_\nu\sigma_e}(\omega_\nu, \omega_e)\big|^2\,+\,\cos^2\theta\sin^2\theta\left[\mathcal{J}^{(1)}_{\sigma_\nu\sigma_e}(\omega_\nu, \omega_e)\,\mathcal{J}^{(2)\hspace{0.2mm}*}_{\sigma_\nu\sigma_e}(\omega_\nu, \omega_e)\,+\,\mathrm{c.c.}\right]\Bigr\}.
\end{eqnarray}
The spin sum for the process $(i)$ is given by
\bea\label{mixspinsum}
&&\frac{1}{T}\sum_{\sigma_\nu,\sigma_e}\left[\mathcal{J}^{(1)}_{\sigma_\nu\sigma_e}(\omega_\nu, \omega_e)\,\mathcal{J}^{(2)\hspace{0.2mm}*}_{\sigma_\nu\sigma_e}(\omega_\nu, \omega_e)\,+\,\mathrm{c.c.}\right]\ =\ \frac{2^3\,\delta(\omega_e-\omega_\nu-\Delta m)}{a^2\,\pi^3\,\sqrt{l_{\nu_1}l_{\nu_2}}}\cosh\left(\pi\omega/a\right)\cosh\left(\pi\omega_e/a\right)\\[2mm]\nonumber
&&\times\,\Biggl[l_e\left(\kappa_\nu^2\,+\,m_{\nu_1}m_{\nu_2}\,+\,l_{\nu_1}l_{\nu_2}\right)\Bigl|K_{i\omega_e/a+1/2}\left(\frac{l_e}{a}\right)\Bigr|^2\mathrm{Re}\left\{K_{i\omega_\nu/a+1/2}\left(\frac{l_{\nu_1}}{a}\right)K_{i\omega_\nu/a-1/2}\left(\frac{l_{\nu_2}}{a}\right)\right\}\\[2mm]\nonumber
&&+\,\Big[\hspace{-0.2mm}\left(k^x_\nu k^x_e\hspace{0.2mm}+\hspace{0.2mm}k^y_\nu k^y_e\right)\left(l_{\nu_1}\hspace{0.2mm}+\hspace{0.2mm}l_{\nu_2}\right)\hspace{0.2mm}+\hspace{0.2mm}m_e\left(l_{\nu_1}m_{\nu_2}\hspace{0.2mm}+\hspace{0.2mm}l_{\nu_2}m_{\nu_1}\right)\hspace{-0.2mm}\Big]\mathrm{Re}\left\{K^2_{i\omega_e/a+1/2}\left(\frac{l_e}{a}\right)K_{i\omega_\nu/a+1/2}\left(\frac{l_{\nu_1}}{a}\right)K_{i\omega_\nu/a+1/2}\left(\frac{l_{\nu_2}}{a}\right)\right\}\Bigg],
\eea
where $\kappa_\nu\equiv(k_\nu^x,k_\nu^y)$.

Next, by performing similar calculations for the other two processes and adding up the three contributions, we finally obtain the total transition rate in the comoving frame:
\begin{equation}
\label{gammacc}
\Gamma_{acc}^{p\rightarrow n}\ =\ 
\cos^4\theta\, \widetilde\Gamma^{p\rightarrow n}_{1}\,+\,\sin^4\theta\,\widetilde\Gamma^{p\rightarrow n}_{2}\,+\,\cos^2\theta\sin^2\theta\,\widetilde\Gamma^{p\rightarrow n}_{12},
\end{equation}
where $\widetilde\Gamma^{p\rightarrow n}_{j}$, $j\,=\,1,2$, is defined as
\begin{equation}
\label{integralbis}
\widetilde\Gamma^{p\rightarrow n}_{j}\ \equiv\ \frac{2\,G_{F}^{2}}{a^2\,\pi^7\,e^{\pi\Delta m/a}}\int_{-\infty}^{+\infty}\hspace{-1.5mm}d\omega\hspace{0.4mm}R_{j}(\omega),\qquad j\,=\,1,2,
\end{equation}
with $\mathcal{R}_j(\omega)$ being defined as in Eq.~(\ref{R}) for each of the two neutrino mass eigenstates, and
\begin{eqnarray}\nonumber
\label{int1212}
\widetilde\Gamma^{p\rightarrow n}_{12}&=&\frac{2\,G_{F}^{2}}{a^2\,\pi^7\,\sqrt{l_{\nu_1}l_{\nu_2}}\,e^{\pi\Delta m/a}}\int_{-\infty}^{+\infty}\hspace{-2mm}d\omega\,\Biggl\{\,\int d^2k_e\,l_e\Bigl|K_{i\omega/a+1/2}\left(\frac{l_e}{a}\right)\Bigr|^2
\int d^2k_\nu\,\big(\kappa_\nu^2\,+\,m_{\nu_1}m_{\nu_2}\,+\,l_{\nu_1}l_{\nu_2}\big)\\[2mm]\nonumber
&&\times\,\mathrm{Re}\left\{K_{i(\omega-\Delta m)/a+1/2}\left(\frac{l_{\nu_1}}{a}\right)K_{i(\omega-\Delta m)/a-1/2}\left(\frac{l_{\nu_2}}{a}\right)\right\}\,+\, m_e\int d^2k_e\int d^2k_\nu\big(l_{\nu_1}m_{\nu_2}\,+\,l_{\nu_2}m_{\nu_1}\big)\\[2mm]
&&\times\,\mathrm{Re}\left\{K^2_{i\omega/a+1/2}\left(\frac{l_e}{a}\right)K_{i(\omega-\Delta m)/a-1/2}\left(\frac{l_{\nu_1}}{a}\right)K_{i(\omega-\Delta m)/a-1/2}\left(\frac{l_{\nu_2}}{a}\right)\right\}\Biggr\}.
\end{eqnarray}

It is now possible to verify that 
\be
\label{equality}
\Gamma^{p\rightarrow n}_{j}\ =	\ \widetilde\Gamma^{p\rightarrow n}_{j}\qquad j\,=\,1,2.
\ee 
By comparing Eqs.~(\ref{eqn:inertresultat}) and~(\ref{gammacc}) and using the above equality, we thus realize that inertial and comoving calculations would match, provided that the integrals Eqs.~(\ref{integral12}) and~(\ref{int1212}) coincide. As in the inertial case, however, the treatment of the $\widetilde\Gamma^{p\rightarrow n}_{12}$ is absolutely non trivial. A clue to a preliminary solution can be found by expanding $\widetilde\Gamma^{p\rightarrow n}_{12}$ in the limit of small neutrino mass difference, as in Sec.~\ref{inerframemix}:
\be\label{approxrind}
\widetilde\Gamma^{p\rightarrow n}_{12}\ =\ 2\widetilde\Gamma^{p\rightarrow n}_1\,+\,\frac{\delta m}{m_{\nu_1}}\,\widetilde\Gamma^{(1)}\,+\, \mathcal{O}\left(\frac{\delta m^2}{m^2_{\nu_1}}\right),
\ee 
where $\widetilde\Gamma^{p\rightarrow n}_1$ is defined in Eq.~(\ref{integralbis}) and we have denoted by $\widetilde\Gamma^{(1)}$ the first-order term of the expansion. For $m_{\nu_1}\rightarrow 0$, it is possible to show that
\be\label{firstordrind}
\frac{\widetilde\Gamma^{(1)}}{m_{\nu_1}}\ =\ \lim_{\varepsilon\rightarrow 0}\hspace{0.3mm}\frac{2^2\,G_F^2\,m_e}{a^2\,\pi^7\,e^{\pi\Delta m/a}}\int_{-\infty}^{+\infty}\hspace{-2mm}d\omega\,\mathrm{Re}\left\{\int d^2k_\nu\hspace{0.3mm} K^2_{i(\omega-\Delta m)/a-1/2}\left(\frac{l_\varepsilon}{a}\right)\int d^2k_e\hspace{0.3mm} K^2_{i\omega/a+1/2}\left(\frac{l_e}{a}\right)\right\},
\ee
where $l_\varepsilon\ =\ \sqrt{(k^x_\nu)^2\,+\,(k^y_\nu)^2\,+\,\varepsilon^2}$, with $\varepsilon$ acting as a regulator. 

Equation~(\ref{firstordrind}) can be now further manipulated by introducing the following relation involving Meijer G-function (see, e.g., Ref.~\cite{Grad}):
\be\label{meijer}
x^{\sigma}K_{\nu}(x)K_{\mu}(x)\ =\ \frac{\sqrt{\pi}}{2}G_{24}^{40}\,\Biggl(x^{2}\Biggr|\begin{matrix}\frac{1}{2}\sigma,\,\frac{1}{2}\sigma\,+\,\frac{1}{2}\\[2mm]\frac{1}{2}(\nu\,+\,\mu\,+\,\sigma),\frac{1}{2}(\nu\,-\,\mu\,+\,\sigma),\frac{1}{2}(-\nu\,+\,\mu\,+\,\sigma),\frac{1}{2}(-\nu\,-\,\mu\,+\,\sigma)
\end{matrix}\Biggr).
\ee 
A somewhat laborious calculation then leads to
\bea\label{bigrind}\nonumber
\frac{\widetilde\Gamma^{(1)}}{m_{\nu_1}}&=&\lim_{\varepsilon\rightarrow 0}\hspace{0.3mm}\frac{2\,G_F^2\,m_e}{a^2\,\pi^4\,e^{\pi\Delta m/a}}\int_{-\infty}^{+\infty}\hspace{-2mm}d\omega\int_{C_s}\frac{ds}{2\pi i}\int_{C_t}\frac{dt}{2\pi i}\int_{0}^{+\infty}\hspace{-2mm}dk_\nu\,k_\nu\,l_\varepsilon^{2s}\int_{0}^{+\infty}\hspace{-2mm}dk_e\,k_e\,l_e^{2t}\\[2mm]\nonumber
&&\times\,\Bigg[\frac{\Gamma\left(-s\right)\Gamma\left(-t\right)\Gamma\left(\frac{i\omega}{a}\hspace{0.2mm}+\hspace{0.2mm}\frac{1}{2}\hspace{0.2mm}-\hspace{0.2mm}t\right)\Gamma\left(-\frac{i\omega}{a}\hspace{0.2mm}-\hspace{0.2mm}\frac{1}{2}\hspace{0.2mm}-\hspace{0.2mm}t\right)\Gamma\left(\frac{i(\omega\hspace{0.2mm}-\hspace{0.2mm}\Delta m)}{a}\hspace{0.2mm}-\hspace{0.2mm}\frac{1}{2}\hspace{0.2mm}-\hspace{0.2mm}s\right)\Gamma\left(-\frac{i(\omega\hspace{0.2mm}-\hspace{0.2mm}\Delta m)}{a}\hspace{0.2mm}+\hspace{0.2mm}\frac{1}{2}\hspace{0.2mm}-\hspace{0.2mm}s\right)}{\Gamma\left(-s\hspace{0.2mm}+\hspace{0.2mm}\frac{1}{2}\right)\Gamma\left(-t\hspace{0.2mm}+\hspace{0.2mm}\frac{1}{2}\right)}\\[2mm]
&&+\,\frac{\Gamma\left(-s\right)\Gamma\left(-t\right)\Gamma\left(\frac{i\omega}{a}\hspace{0.2mm}-\hspace{0.2mm}\frac{1}{2}\hspace{0.2mm}-\hspace{0.2mm}t\right)\Gamma\left(-\frac{i\omega}{a}\hspace{0.2mm}+\hspace{0.2mm}\frac{1}{2}\hspace{0.2mm}-\hspace{0.2mm}t\right)\Gamma\left(\frac{i(\omega\hspace{0.2mm}-\hspace{0.2mm}\Delta m)}{a}\hspace{0.2mm}+\hspace{0.2mm}\frac{1}{2}\hspace{0.2mm}-\hspace{0.2mm}s\right)\Gamma\left(-\frac{i(\omega\hspace{0.2mm}-\hspace{0.2mm}\Delta m)}{a}\hspace{0.2mm}-\hspace{0.2mm}\frac{1}{2}\hspace{0.2mm}-\hspace{0.2mm}s\right)}{\Gamma\left(-s\hspace{0.2mm}+\hspace{0.2mm}\frac{1}{2}\right)\Gamma\left(-t\hspace{0.2mm}+\hspace{0.2mm}\frac{1}{2}\right)}\Bigg].
\eea
In order to perform the integration with respect to $\omega$, let us use the first Barnes lemma, according to which~\cite{Grad}
\be\label{barnes}
\int_{-i\infty}^{+i\infty}\hspace{-2mm}d{\omega}\,\Gamma(a\,+\,{\omega})\Gamma(b\,+\,{\omega})\Gamma(c\,-\,{\omega})\Gamma(d\,-\,{\omega})\ =\ 2\pi i\,\,\frac{\Gamma(a\,+\,c)\Gamma(a\,+\,d)\Gamma(b\,+\,c)\Gamma(b\,+\,d)}{\Gamma(a\,+\,b\,+\,c\,+\,d)}.
\ee
Inserting this relation into Eq.~(\ref{bigrind}), it follows that
\bea
\label{lastacc}
\frac{\widetilde\Gamma^{(1)}}{m_{\nu_1}}&=&\lim_{\varepsilon\rightarrow 0}\frac{G_F^2\,m_e\,a^3}{\pi^3\,e^{\pi\Delta m/a}}\int_{C_s}\frac{ds}{2\pi i}\int_{C_t}\frac{dt}{2\pi i}\left(\frac{\varepsilon}{a}\right)^{2s+2}\left(\frac{m_e}{a}\right)^{2t+2}\frac{\Gamma(-2s)\Gamma(-2t)\Gamma(-t-1)\Gamma(-s-1)}{\Gamma(-s+\frac{1}{2})\Gamma(-t+\frac{1}{2})\Gamma(-2s-2t)}\\[2mm]\nonumber
&&\times\left[\Gamma\left(-s\hspace{0.2mm}-\hspace{0.2mm}t\hspace{0.2mm}+\hspace{0.2mm}1\hspace{0.2mm}+\hspace{0.2mm}i\hspace{0.2mm}\frac{\Delta m}{a}\right)\Gamma\left(-s\hspace{0.2mm}-\hspace{0.2mm}t\hspace{0.2mm}-\hspace{0.2mm}1\hspace{0.2mm}-\hspace{0.2mm}i\hspace{0.2mm}\frac{\Delta m}{a}\right)\,+\,\Gamma\left(-s\hspace{0.2mm}-\hspace{0.2mm}t\hspace{0.2mm}+\hspace{0.2mm}1\hspace{0.2mm}-\hspace{0.2mm}i\hspace{0.2mm}\frac{\Delta m}{a}\right)\Gamma\left(-s\hspace{0.2mm}-\hspace{0.2mm}t\hspace{0.2mm}-\hspace{0.2mm}1\hspace{0.2mm}+\hspace{0.2mm}i\hspace{0.2mm}\frac{\Delta m}{a}\right)\right]\hspace{-0.5mm},
\eea
which is exactly the same expression obtained in the inertial frame, Eq.~(\ref{lastin}).

\section{Conclusions}
In the present paper we have discussed the decay of uniformly accelerated protons. Following the line of reasoning of Refs.~\cite{Matsas:1999jx3,Suzuki:2002xg}, we have reviewed the calculation of the total decay rate both in the laboratory and comoving frame, highlighting the incompatibility between the two results when taking into account neutrino flavor mixing~\cite{Aluw}. Such an inconsistency would not be striking if the underlying theory were not generally covariant, but this is not the case, since the fundamental ingredients for analyzing the process, namely the SM and QFT in curved space-time, are by construction generally covariant. On the other hand, the authors of Ref.~\cite{Aluw} argue their result claiming that mixed neutrinos are not
representations of the Lorentz group with a well-defined
invariant $P^2$, and that the mathematical
origin of the disagreement arises from the noncommutativity of weak and energy-momentum currents. 
Furthermore, they propose the experimental investigation as the only way to resolve
such a controversial issue. Even assuming there are no flaws in this reasoning, we believe the last statement to be basically incorrect: an experiment, indeed, should not be used as a tool for checking the internal consistency of theory against a theoretical paradox.

Led by these considerations, we have thus revised calculations of Ref.~\cite{Aluw} modifying some of the key assumptions of that work. In particular, we have required the asymptotic neutrino states to be flavor rather than mass eigenstates. Within this framework, by comparing the obtained expressions for the two decay rates, it has been shown that they would coincide, provided that the off-diagonal terms Eqs.~(\ref{integral12}) and~(\ref{int1212}) are equal to each other. In order to check whether this is the case, we have performed the reasonable approximation of small neutrino mass difference, pushing our analysis up to the first order in $\frac{\delta m}{m_{\nu}}$. However, due to computational difficulties, the further assumption of vanishing neutrino mass $m_{\nu_1}\rightarrow 0$ has proved to be necessary for getting information about these terms. In such a regime, we have found that Eqs.~(\ref{lastin}) and~(\ref{lastacc}) are perfectly in agreement,  thus removing the aforementioned ambiguity at a purely theoretical level.

Relying on this, one can state that the theoretical framework underlying our result is the correct one in the context of neutrino mixing, provided flavor neutrinos are considered to be the fundamental objects populating the thermal FDU state. Further aspects of the fascinating problem firstly addressed in Ref.~\cite{Aluw} can be investigated only when an exact evaluation of the two decay rates will be available: work is in progress along this direction~\cite{In prepa}, also in view of the non-trivial nature 
of neutrino flavor states in QFT~\cite{Blasone:1995zc}.

\section*{Acknowledgements}
We thank Luca Smaldone for useful discussions on some issues contained in this paper.

\appendix*
\section{Evaluation of the  decay rate in the comoving frame using mass eigenstates}
For the purpose of comparison with the results of Ref.~\cite{Aluw}, in this Appendix we perform explicitly the calculation of the decay rate in the comoving frame using neutrino mass eigenstates as asymptotic  states. Such a choice has been adopted in Ref.~\cite{Aluw} in order to preserve the thermal KMS condition for the accelerated neutrino vacuum. As done in Sec.~\ref{prdec}, we shall limit ourselves to analyze the process $(i)$ in Eq.~(\ref{threeprocesses}). Following Ref.~\cite{Aluw}, it can be shown that the differential transition rate takes the form
\begin{eqnarray}
\label{difmixrind}
\frac{1}{T}\frac{d^{6}\mathcal{P}^{p\rightarrow n}_{(i)}}{d\omega_{\nu}\hspace{0.2mm}d\omega_{e}\hspace{0.2mm}d^2k_{\nu}\hspace{0.2mm}d^2k_e}\,=\,\frac{1}{T}\frac{\hspace{0.2mm}G_{F}^{2}}{2^6\pi^{4}}\hspace{0.2mm}\frac{\sum_{\sigma_{\nu},\sigma_e}\left[\cos^2\theta\,|\mathcal{J}^{(1)}_{\sigma_\nu\sigma_e}(\omega_\nu, \omega_e)|^2\,+\,\sin^2\theta\,|\mathcal{J}^{(2)}_{\sigma_\nu\sigma_e}(\omega_\nu, \omega_e)|^2\right]}{e^{\pi\Delta m/a}\cosh(\pi\omega_{\nu}/a)\cosh(\pi\omega_{e}/a)}.
\end{eqnarray}
where $\mathcal{J}^{(j)}_{\sigma_\nu\sigma_e}(\omega_\nu, \omega_e)$, $j\,=\,1,2$, is defined as in Eq.~(\ref{eqn:j}) for each of the two neutrino mass eigenstates. By performing similar calculations for the processes $(ii)$ and $(iii)$ and adding up the three contributions, we thus obtain
\begin{equation}
\label{accresult}
\Gamma_{acc}^{p\rightarrow n}\ \equiv\ \Gamma_{(i)}^{p\rightarrow n}\,+\,\Gamma_{(ii)}^{p\rightarrow n}\,+\,\Gamma_{(iii)}^{p\rightarrow n}
\ =\ \cos^2\theta\, \widetilde\Gamma^{p\rightarrow n}_{1}\,+\,\sin^2\theta\,\widetilde\Gamma^{p\rightarrow n}_{2},
\end{equation}
where  we have used the shorthand notation Eq.~(\ref{integralbis}).

By comparing Eqs.~(\ref{eqn:inertresultat}) and~(\ref{accresult}), it is clear that the two decay rates are not in agreement with each other\,--\,a result that is incompatible with the General Covariance of the underlying formalism.
As long as the asymptotic neutrino states in the comoving frame are assumed to be mass eigenstates, however, such a contradiction cannot be resolved, as it is evident for at least two reasons. First, in Eq.~(\ref{accresult}) there is no counterpart of the off-diagonal term found in Eq.~(\ref{eqn:inertresultat}). Furthermore, although the partial decay rates Eqs.~(\ref{integral}) and~(\ref{integralbis}) are verified to coincide with each other (see Eq.~(\ref{equality})), Pontecorvo matrix elements appear in Eqs.~(\ref{eqn:inertresultat}) and~(\ref{accresult}) with different powers, thereby also preventing the identification of the diagonal terms.

\end{document}